\documentclass[aps,prl,10pt,twocolumn,superscriptaddress]{revtex4}
\usepackage[bbgreekl]{mathbbol}
\usepackage{amsmath}
\usepackage{amssymb}
\usepackage{bbm}
\usepackage{graphicx}
\usepackage{framed}
\usepackage{xcolor}
\usepackage{hyperref}
\usepackage{epstopdf}
\usepackage{dsfont}
\usepackage{amssymb}
\usepackage{amsmath}
\usepackage{amsthm}
\usepackage{wrapfig}
\usepackage{relsize}
\usepackage{bm}
\usepackage{hyperref}

\usepackage{enumitem}
\usepackage{natbib}
\usepackage{graphicx}
\usepackage{physics}
\usepackage{hyperref}
\usepackage{subfigure}

\begin{document}

	\title{Estimating Quantum and Private capacities of Gaussian channels via degradable extensions}
	
	\author{Marco Fanizza}
	\email{marco.fanizza@sns.it}
	\author{Farzad Kianvash}
	\email{farzad.kianvash@sns.it}
	\author{Vittorio Giovannetti}	\affiliation{NEST, Scuola Normale Superiore and Istituto Nanoscienze-CNR, I-56126 Pisa, Italy}
	\date{\today}

	\begin{abstract}
		We present upper bounds on the quantum and private capacity of single-mode, phase-insentitive Bosonic Gaussian Channels  
				 based on degradable extensions. Our 
		findings are  state-of-the-art in the following parameter regions: low temperature and high transmissivity for the thermal attenuator, low temperature  for additive Gaussian noise, high temperature and intermediate amplification for the thermal amplifier.
	 \end{abstract}
	
	\maketitle

\textit{Introduction}.-- Quantum Shannon theory \cite{HOL BOOK, WILDE BOOK} provides a characterization of the maximum achievable transmission rates (capacities) for classical or quantum data through a quantum channel, as maximizations of entropic functionals. Available characterizations of most capacities cannot be computed algorithmically, since they involve a limit of an infinite sequence of optimization problems, one for any number of uses of the channel. Superadditivity of quantum entropic functionals makes such regularization necessary~\cite{shor superadditivity DC, Di vincenzo superadditivity DC, smith superadditivity, Fern superadditivity, q superadditivity, gaussian q superadditivity, Cubit superadditivity, dephrasure, bl superadditivity,c superadditivity,p superadditivity,tradeoff1,tradeoff2, Siddhu}, and can hinder the evaluation of capacities even for simple fundamental channels. While it is hard to get past regularized expressions in the general case, it is important to improve our best understanding of the capacities of channels of physical interest. Here we focus on quantum and private capacities, which give the optimal transmission rates of quantum states and private classical information over a quantum channel~\cite{S,Lioyd,Q CAP DEV,privCWY,privDS} and are characterized as optimization of the coherent information and the private information, respectively. We consider realistic single-mode noise models for communication across free space or optical fiber: thermal attenuators, thermal amplifiers and additive Gaussian noise~\cite{CAVES,REV1,BOSrew1,serafini}.
These channels collectively known as phase-insensitive, single-mode Bosonic Gaussian Channels, can be understood as unitary interactions with a single-mode thermal environment mediated by Hamiltonians which are quadratic in the operators fields of the model. The quantum and private capacities of these type of channels have been extensively studied~\cite{swat, Pirandola upp bound, btw, Rosati, Noh2018, Noh2020, Lim2019}, but they are still unknown. Interesting variations are Gaussian interaction with general environment~\cite{Lim2019a, Lami2020}, and environmental assisted communication \cite{Oskouei2021}. 

In this work we find new state-of-the-art upper bounds on the quantum and private capacity of thermal attenuators, amplifiers and additive Gaussian noise, using degradable extensions. For degradable channels coherent and private information are additive, and one can compute quantum and private capacities using the same single letter formula~\cite{privDS,deg channels,Smithpdeg}. Therefore, an established strategy to find upper bounds for quantum and private capacity is to exploit (approximately) degradable channels which reduce to the channel of interest after pre and/or post-processing \cite{SS UPP B DEP,Ouyang,SUTT UPP B DEP, lownoiseQ, distdepo,Rosati, swat, flagged channel 1,flagged channel 2,wang}, and use the capacity of the extended channels as an upper bound.

To upper bound the quantum capacity of the additive Gaussian noise, we introduce a flagged extension of the additive Gaussian noise which is degradable. A flagged extension of a convex combination of Completely Positive and Trace Preserving (CPTP) maps~\cite{HOL BOOK,WILDE BOOK} 
 $\Lambda[\rho]=\sum_i p_i \Lambda_i[\rho]$ can be written as $\Lambda^{e}[\rho]=\sum_i p_i \Lambda_i[\rho]\otimes \sigma_i$, such that an auxiliary system encodes into the states $\sigma_i$ information on which noise model is tampering the communication line. In discrete variable quantum information, degradable flagged extensions offer the best known upper bounds to several important channels~\cite{wang, flagged channel 2}. In this article we extend this approach to continuous variable systems. In particular, we are inspired by the sufficient conditions of degradability of finite-dimensional flagged channels introduced in~\cite{flagged channel 2}. The new upper bound is state-of-the-art in the low temperature regime. Using this upper bound and data-processing, we also bound the quantum capacity of thermal amplifier, obtaining a state-of-the-art bound in the high temperature regime, for intermediate amplification values. 
We also improve the construction in \cite{Rosati} based on weak degradability to find a noisier degradable extension of the thermal attenuator, and show that it gives the best upper bound in the low noise regime, for low temperature and high transmissivity. 

In the following, we first review basic facts about quantum and private capacities and the formalism of Gaussian states and channels. Next, we present the flagged extension of the additive Gaussian noise and the extension of the thermal attenuator. In both sections, we compute their quantum and private capacities, and compare the new available upper bounds with previous results. 

\textit{Preliminaries}.-- For a possibly infinite dimensional Hilbert space $\mathcal{H}$, we can define the set of all trace-class linear operators as $\mathcal{T(H)}$.
A quantum channel $\Lambda:\mathcal{T}(\mathcal{H}_A)\rightarrow\mathcal{T}(\mathcal{H}_B)$ is a completely positive trace preserving (CPTP) linear map on the space of trace-class operators. Quantum states $\mathcal{G(H)}$ are identified with positive semi-definite operators with trace one. Any quantum channel can be represented in Stinespring representation
as $\Lambda[\hat \theta]=\Tr_E (\hat V\hat \theta {\hat V}^\dagger)\,$, where $\hat V:\mathcal{H}_A\rightarrow\mathcal{H}_{BE}$ is an isometry. A complementary channel $\tilde{\Lambda}:\mathcal{T}(\mathcal{H}_A)\rightarrow\mathcal{T}(\mathcal{H}_E)$ is defined as following $\Lambda^{c}[\hat \theta]:=\Tr_A [\hat V\hat \theta {\hat V}^\dagger]$. If there exists another channel $W:\mathcal{T}(\mathcal{H}_B)\rightarrow\mathcal{T}(\mathcal{H}_E)$ satisfying $W\circ\Lambda=\Lambda^{c}$, $\Lambda$ is said to be degradable. 
The quantum capacity of a quantum channel is the maximum rate at which it can transmit quantum information reliably over asymptotically many uses of the channel. It is equal to the regularized coherent information, i.e.,
\begin{equation}
Q(\Lambda) = \lim_{n\to\infty} 
  {Q^{(1)}(\Lambda^{\otimes n})}/{n},
  \end{equation}
where $Q^{(1)}(\Lambda):= \sup_{\hat \rho \in \mathcal{G(H}_A)} S(\Lambda(\hat\rho)) - S(\Lambda^c(\hat\rho))$ is the coherent information of $\Lambda$,
 $\Lambda^c$ is a complementary channel of $\Lambda$, and $S(\hat\rho) := - \mbox{Tr}[ \hat\rho \log_2 \hat\rho]$ is the von~Neumann entropy of the state $\hat\rho$.
For degradable channels the coherent information is additive \cite{privDS,deg channels} and it holds $Q(\Lambda)=Q^{(1)}(\Lambda)$.
The private capacity is characterized by a different regularized expression, but it coincides with the quantum capacity for degradable channels~\cite{Smithpdeg}. Thus, since the upper bounds we obtain in this work come from degradable extensions, they are immediately understood as upper bounds for the private capacity too.

In this work we consider infinite dimensional Hilbert spaces $L^2(\mathbb R^n)$ of square integrable functions, corresponding to $m$ modes of harmonic oscillators chacterized by position and momentum operators which we group as $\mathbf{\hat{r}}=(\hat{x}_1,\hat{p}_1,...,\hat{x}_n,\hat{p}_n)^\text{T}$. They   satisfy the canonical commutation relations $[\hat{r}_i,\hat{r}_j]=i\Omega_{ij} I$, where
 \begin{equation}
 	 \Omega:=\left(\begin{matrix}0&1\\-1&0\end{matrix}\right)^{\oplus n}.
 \end{equation}
 For  ${\mathbf{r}}\in {\mathbb R}^{2n}$
 we introduce the (unitary) displacement operators $\hat D_{\mathbf{r}}=e^{i {\mathbf{r}}^{\sf{T}}\Omega \mathbf{\hat{r}}}$ satisfying the identity $\hat D_{\mathbf{r_1+r_2}}=\hat D_{\mathbf{r_1}}\hat D_{\mathbf{r_2}}e^{i {\mathbf{r}_1}^{\sf{T}}\Omega {\mathbf{r}}_2/2}$, and define the characteristic function of a trace class operator $\hat\rho$ on $L^{2}(\mathbb R^n)$ as
$\chi(\mathbf r):=\Tr[\hat \rho \hat D_{-\mathbf{r}} ]$~\cite{serafini,BOSrew1}.
The Gaussian states are those states such that their characteristic function is Gaussian, i.e.
$
\chi(\mathbf r)=\exp{-\frac{1}{4}{\mathbf r}^{\sf T}\Omega^{\sf T} V\Omega {\mathbf r} +i{\mathbf r}^{\sf T}\Omega m}$, 
with $\mathbf{m}:=\Tr[\mathbf{\hat{r}}\hat\rho]$ and $V:=\Tr[\{\mathbf{(\hat{r}-m),(\hat{r}-m)}^\text{T}\}\hat\rho]$ being the associated statistical mean vector and covariance matrix where $\{\hat{A},\hat{B}\}:=\hat{A}\hat{B}+\hat{B}\hat{A}$.
Bosonic Gaussian Channels can now be identified with the CPTP super-operators on $n$ modes that map the set of Gaussian states to itself~\cite{REV1,BOSrew1,serafini} and  can hence be charaterized by how they transform $\mathbf m$ and $V$.
In the following we shall specifically focus on single mode ($n=1$) 
 thermal attenuator BGCs $\mathcal{E}_{\eta,N}$ defined by the mapping~\cite{CARUSO} 
 \begin{align}
&\mathbf{m}\xrightarrow{\mathcal{E}_{\eta,N}} \mathbf{m}'=\sqrt{\eta}{\bf m}\;,\\
&V\xrightarrow{\mathcal{E}_{\eta,N}}V'=\eta V + (1-\eta)(2N+1)I_2\,  ,
\end{align}
with $0\leq\eta\leq 1$ and $N\geq 0$ being the characteristic parameters of the model, and 
$I_2$ being the  two dimensional identity.
 Via Stinespring representation~\cite{REV1,BOSrew1,serafini}
 these special transformations are better understood in terms of a beam splitter coupling
 with an extra  an enviromental mode $E$ initialized in a thermal (Gaussian) state.
 Specifically labelling with $A$ the system mode, and indicating with $\hat\rho_A$  its input state we
 can write  \begin{equation}\label{stinextatt}
\mathcal{E}_{\eta,N}[\hat\rho_A]:=\Tr_{E}[ \hat U_{\eta}(\hat\rho_{A}\otimes\hat \tau_{E})].
\end{equation}
In this expression  $\Tr_{E}$ represents the partial trace over the environment while 
 $\hat U_\eta$ ($0\leq\eta\leq 1$)  is two-mode unitary operator that transforms $\mathbf{\hat{r}}$ according to
 \begin{equation}
 	\hat U_\eta \mathbf{\hat{r}} \hat U^\dagger_\eta=\begin{pmatrix}
 		\sqrt{\eta}I_2 & \sqrt{1-\eta}I_2 \\
 		-\sqrt{1-\eta}I_2 & \sqrt{\eta}I_2
 		\end{pmatrix} \mathbf{\hat{r}}. 
 \end{equation}
 The thermal state $\hat\tau$ entering in (\ref{stinextatt}) is finally defined by $\mathbf{m}_{\hat\tau}=(0,0)$ and $V_{\hat \tau}=(2N+1)I_2$. Such density matrix  is purified by  a two-mode squeezed state $\ket{\tau}$, which has $\mathbf{m}_{\ket{\tau}}=(0,0,0,0)$ and
 \begin{equation}
  V_{\ket{\tau}}=\left(\begin{matrix} (2N+1)V_{\hat \tau} & \sqrt{V_{\hat \tau}^2-1}\sigma_3 \\ \sqrt{V_{\hat \tau}^2-1}\sigma_3  &(2N+1)V_{\hat \tau}\end{matrix}\right),\quad \sigma_3=\left(\begin{matrix}1&0\\0&-1\end{matrix}\right).
 \end{equation}
We shall also consider single-mode thermal amplifier BGCs ${\Phi}_{g,N}$. In this case the input state interacts with a thermal bath through a two mode squeezing operator with parameter $g\geq 1$ inducing the mapping{ \begin{align}
&\mathbf{m}\xrightarrow{{\Phi}_{g,N}} \mathbf{m}'=\sqrt{g}\mathbf{m}\;, \\
&V\xrightarrow{{\Phi}_{g,N}}V'=g V + (g-1)(2N+1)I_2\,  .
\end{align}
Finally we study the single-mode Additive Gaussian Noise  Channel (AGNC) $\Lambda_\beta$  that can be expressed as 
\begin{equation}
\Lambda_\beta[\hat\rho]:=\frac{\beta}{2\pi}\int_{\mathbb R^{2}} \mathrm{d} {\mathbf{r}} e^{-\frac{\beta}{2}\mathbf{r}^{\sf T}\mathbf{r}} \hat D_{\mathbf{r}}\hat\rho \hat D_{\mathbf{r}}^\dagger ,  
\end{equation} 
where $\beta>0$ is the inverse temperature, that induces the mapping 
\begin{align}
&\mathbf{m}\xrightarrow{{\Lambda}_{\beta}} \mathbf{m}'=\mathbf{m}\;,\\
&V\xrightarrow{{\Lambda}_{\beta}}V'=V + {2} I_2/\beta\,  .
\end{align}
	
{\textit{Upper bounds for the AGNC}.--  In the high temperature regime (i.e. for $1/\beta\geq 0.5$)
 the quantum capacity of the channel  $\Lambda_{\beta}$ is known to be exactly zero, using data-processing techniques~\cite{HOL BOOK}. {In particular in~\cite{Noh2018}, the following upper bound is computed using data-processing:
\begin{equation}\label{NAJ}
Q(\Lambda_{\beta})\leq Q_{\text{NAJ}}(\beta)=\max\{\log_2(\beta-1),0\},
\end{equation}

and correctly gives $Q(\Lambda_{\beta})=0$ for $1/\beta\geq 0.5$.}
 In the low temperature instead we only have a lower bound for 
  $Q(\Lambda_{\beta})$ {given by the coherent information for one use of the channel, evaluated on an infinite temperature state,} i.e. 
\begin{equation}\label{q1 agn}
Q(\Lambda_{\beta}) \geq {Q_L(\Lambda_{\beta}) }:=\max\{ \log_2\beta-1/\ln 2, 0\}\;, 
\end{equation}
and the inequality,
\begin{equation}\label{PLOB} 
Q(\Lambda_{\beta}) \leq Q_{\mathrm{PLOB}}(\beta)=\log_2 \beta -1/\ln 2 +{1}/{(\beta\ln 2)}\;,
\end{equation}
which follows from  an upper bound  for the generic two-way quantum capacity~\cite{Pirandola upp bound}. 

Generalizing  the procedure introduced in~\cite{flagged channel 2} we present an improved upper bound
considering the following flagged extension of  $\Lambda_{\beta}$, i.e. 
  \begin{align}
\Lambda^{e}_{\beta}[\hat\rho]
&:=\frac{\beta}{2\pi} \int_{\mathbb R^2}   \mathrm{d} {\mathbf{r}} e^{-\frac{\beta}{2}\mathbf{r}^{\sf T}\mathbf{r}} \hat D_{\mathbf{r}}\hat\rho \hat D_{\mathbf{r}}^\dagger \otimes\ketbra{\phi_\mathbf{r}}\;,
\end{align}
where setting $\mathbf{r}:=(x,p)$,  $\ket{\phi_\mathbf{r}}$ are product of displaced squeezed states defined by 
\begin{equation} \label{flagsNEW} 
\ket{\phi_\mathbf{r}}:=\hat{D}_{(0,-p/2)}\ket{\beta/2}\otimes \hat{D}_{(0,x/2)}\ket{\beta/2}\;, 
\end{equation}
with  $\ket{\beta/2}$ being a single-mode squeezed vacuum with mean values $\mathbf{m}=0$ and covariance matrix $V=\tiny{\left(\begin{matrix}2/\beta &0 \\ 0 & \beta/2\end{matrix}\right)}$. 
As explicitly shown in~\cite{SM}   $\Lambda^{e}_{\beta}$ 
 is degradable: here we notice that  
 the intuition for choosing  the flags states as in~(\ref{flagsNEW}) comes from a result by the same authors on finite dimensional channels \cite{flagged channel 2}. Indeed there we proved that if $\{\sqrt{p_i}\hat{U}_i\}_{i=1,...,n}$ are unitary Kraus operators of a CPTP map $\mathcal N$ and $\{\ket{i}\}_{i=1,...,n}$ is an orthonormal basis for flags, sufficient conditions for the degradability of ${\mathcal N^e}[\hat{\rho}]:=\sum_{i=1}^{n} p_i \hat{U}_i \hat{\rho} \hat{U}_i^\dagger \otimes \ket{\phi_i}\bra{\phi_i}$ are the following
\begin{equation}
		\bra{i'}\ket{\phi_i}\sqrt{p_i}\hat{U}_{i'} \hat{U}_i=\bra{i}\ket{\phi_{i'}}\sqrt{p_{i'}}\hat{U}_{i} \hat{U}_{i'}\quad \forall i,i' \, .
\end{equation}
If we can use a continuous set of flags, replacing the orthonormal basis for the flag space withc the (two-mode) pseudo-eigenbasis $\{ |\rm{x}_1,\rm{x}_2\rangle\}_{\rm{x}_1,\rm{x}_2\in {\mathbb R}}$ of the position operators of the ancillary modes, applying this result to $\Lambda^{e}_{\beta}$ we get
\begin{align}
&\bra{\gamma x',\gamma p'}\ket{\phi_\mathbf{r}}e^{-\frac{\beta}{4}\mathbf{r}^{\sf T}\mathbf{r}}\hat D_{\mathbf{r}'} \hat D_{\mathbf{r}}{=}\bra{\gamma x,\gamma p}\ket{\phi_{\mathbf{r}'}}e^{-\frac{\beta}{4}{\mathbf{r}'}^{\sf T}\mathbf{r}'}\hat D_{\mathbf{r}}\hat D_{\mathbf{r}'} \, , \label{questa} 
\end{align}
for all $\mathbf{r}'=(x',p'), \mathbf{r}=(x,p) \in {\mathbb R}^2$ with $\gamma$ a suitable rescaling factor to determine. 
 With the choice~(\ref{flagsNEW})  of the flags we have explictly 
 \begin{equation}\label{flags}
\bra{\gamma x',\gamma p'}\ket{\phi_\mathbf{r}}=\sqrt{\frac{\beta}{2\pi }}e^{-\beta\frac{\gamma^2x'^2+\gamma^2p'^2}{4}-\gamma \frac{ip'x-ix'p}{2}},
\end{equation}
which satisfies the condition~(\ref{questa}) with $\gamma=1$. 
Exploiting the degradability of 
 $\Lambda^e_{\beta}$ and the fact that it is gauge-covariant (in a generalized sense that we specify in~\cite{SM}), the capacity of this map can 
 be easily computed~\cite{HOL BOOK} leading to the following inequality 
\begin{equation}\label{upper b flags}
Q(\Lambda_{\beta}) \leq Q(\Lambda^e_{\beta})=\log_2 \beta-1/\ln 2+2h \left({\sqrt{1+1/\beta^2}}\right)\;, 
\end{equation}
with $h(x):=\frac{x+1}{2}\log_2\left(\frac{x+1}{2}\right)-\frac{x-1}{2}\log_2\left(\frac{x-1}{2}\right)$
(see~\cite{SM} for details).
As shown in Fig.~\ref{fig:addnoisecomp}, 
Eq.~(\ref{upper b flags}) is better than~(\ref{PLOB})  where $Q(\Lambda_{\beta})$ is supposed to be non-zero, i.e. 
for  $1/\beta\leq 0.5$. 

\textit{Upper bounds for the thermal amplifier}.-- 
Invoking data-processing inequality~\cite{WILDE BOOK}, we can use 
 Eq.~(\ref{upper b flags}) to upper bound 
the quantum capacity  thermal amplifier ${\Phi}_{g,N}$ (a similar argument could also
be invoked for the 
 thermal attenuator $\mathcal{E}_{\eta,N}$ but the result we get is worst than 
 the bounds reported in the next section). As a matter of fact 
 any thermal amplifier can be written as a composition of a zero temperature amplifier and an AGNC i.e. 
	$\Phi_{g,N}=\Lambda_{\tilde{\beta}}\circ\Phi_{g,0}$
with $\tilde{\beta}=\frac{1}{(g-1)N}$. Therefore we get 
\begin{align}
Q(\Phi_{g,N})&\leq Q(\Lambda_{\tilde{\beta}})\leq Q(\Lambda^e_{\tilde{\beta}})\;, \label{newbound} \\ 
Q(\Phi_{g,N})&\leq Q(\Lambda_{\tilde{\beta}})\leq Q_{\mathrm{NAJ}}(\tilde{\beta})\label{addnoiseNAJ}\;, 
\end{align}
{The bound of Eq.~(\ref{addnoiseNAJ}) comes directly to the data-processing decompositions, but to our knowledge it was not explicitly pointed out previously, being implicit in Proposition 12.65 of~\cite{HOL BOOK}. It is the best bound at high $g$ for any $N$, and it gives zero quantum capacity for the lowest $g$. The new bound in Eq.~(\ref{newbound}), in the high temperature regime $N>5$ and for intermediate values of $g$ is provably better than 
the previous best bound reported in~\cite{Pirandola upp bound}}:
\begin{equation}\label{AMPLOB} 
Q(\Phi_{g,N})\leq Q_{\mathrm{AmPLOB}}(g,N):=\log_2(\tfrac{g^{N+1}}{g-1})-h(2N+1)\, .
\end{equation} 
A comparison between these function is reported in Fig.~\ref{fig:ampcomp}, together with the lower bound {given by the coherent information for one use of the channel, evaluated on an infinite temperature state,}
\begin{equation}\label{amplow}
	Q(\Phi_{g,N}) \geq {Q_L}(\Phi_{g,N}):=\max\{\log_2(\tfrac{g}{g-1})-h(2N+1),0\}\;. 
\end{equation}

\begin{figure}
	\centering
	\includegraphics[width=1.0\linewidth]{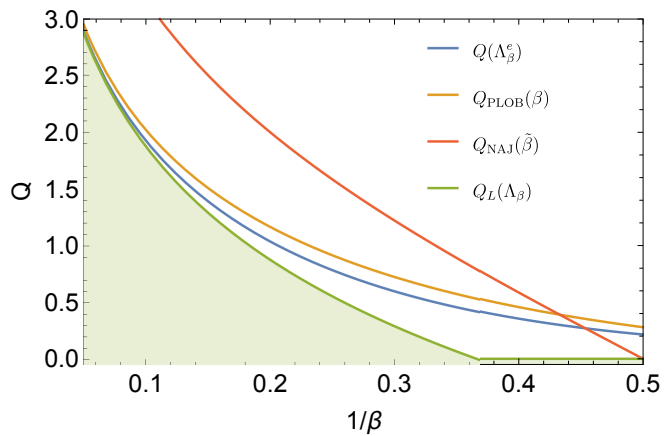}
	\caption{Quantum capacity region for  the AGNC $\Lambda_{\beta}$: 
	comparison of the upper bound $Q(\Lambda^e_{\beta})$ of Eq.~(\ref{upper b flags}) with $Q_{\mathrm{PLOB}}$ in Eq.~(\ref{PLOB})~\cite{Pirandola upp bound} and $Q_{\text{NAJ}}(\beta)$ in Eq.~(\ref{NAJ})~\cite{Noh2018}. $Q_L(\Lambda_\beta)$ is the lower bound of Eq.~(\ref{q1 agn}).}
	\label{fig:addnoisecomp}
\end{figure}
\begin{figure}
\centering
	\includegraphics[width=1.0\linewidth]{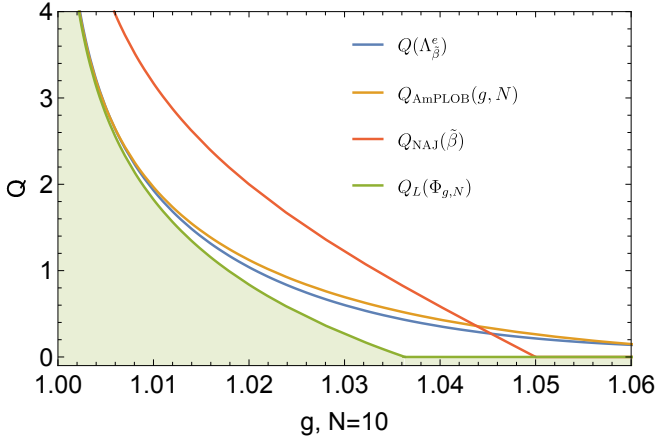}
	\caption{Quantum capacity region for  the thermal amplifier channel $\Phi_{g,N}$: 
	comparison between the upper bound $Q(\Lambda^e_{\tilde{\beta}})$ of Eq.~(\ref{newbound}) with $Q_{\mathrm{AmPLOB}}(g,N)$ of Eq.~(\ref{AMPLOB})~\cite{Pirandola upp bound} and $Q_{\text{NAJ}}(\tilde\beta)$ in Eq.~(\ref{NAJ})~\cite{Noh2018} for $N=10$. $Q_L(\Phi_{g,N})$ is the lower bound of Eq ~(\ref{amplow})}
	\label{fig:ampcomp}
\end{figure}

\textit{Upper bounds for the thermal attenuator}.--  To deal with the quantum capacity of the channel $\mathcal{E}_{\eta,N}$ of Eq.~(\ref{stinextatt})  we construct 
 a degradable  extension of such map. 
We first define the passive unitary operator
\begin{equation}
\hat W_{\eta}:={\hat U {}_{\eta}}_{AE}\otimes {\hat U{}_{\eta}}_{A'E'}\;, 
\end{equation}
where ${\hat U{}_{\eta} }_{AE}$ and ${\hat U{}_{\eta}}_{A'E'}$ are beam splitters transformations acting respectively on the pair of modes $A,E$ and $A',E'$. 
We introduce hence the channel 
\begin{equation}\label{stinextattext}
\mathcal F_{\eta,N}[\rho_{AA'}]:=\Tr_{EE'}[\hat W_{\eta}(\rho_{AA'}\otimes\ketbra{\tau}_{EE'})\hat W_{\eta}^{\dagger}], 
\end{equation}
and define the extension of $\mathcal{E}_{\eta,N}$ as
\begin{equation}\label{extatt}
\mathcal{E}^{e}_{\eta,N}[\rho_A]:=\mathcal F_{\eta,N}(\rho_{A}\otimes\ketbra{0}_{A'}).
\end{equation}
The map $\mathcal F_{\eta,N}$ is manifestly Gaussian, and its action on the first and second moments is
\begin{align}
\bf{m}&\xrightarrow{{\mathcal{F}}_{\eta,N}} {\bf{m}}'=\sqrt{\eta}{\bf m}\;,\\
V&\xrightarrow{{\mathcal{F}}_{\eta,N}} V' =\eta V+(1-\eta)V_{\ket{\tau}}\;,
\end{align}
With the Stinespring representation in Eq. (\ref{stinextattext}) the complementary channel can now computed as $\mathcal{F}^{c}_{\eta,N}=\mathcal{F}_{1-\eta,N}$. Simple algebra shows that if $\eta>1/2$ then 
  \begin{equation}
   {\mathcal{F}}^{c}_{\eta,N}=\mathcal{F}_{1-\eta,N}=  {\mathcal{F}}_{(1-\eta)/\eta,N}\circ {\mathcal{F}}_{\eta,N}\;, 
\end{equation}
implying that in such regime 
 ${\mathcal{F}}_{\eta,N}$ (and thus ${\mathcal{E}}^{e}_{\eta,N}$)  is degradable.   
The quantum capacity of ${\mathcal{E}}^{e}_{\eta,N}$ can be thus calculated by evaluating the coherent information of the channel
leading to
\begin{eqnarray}  \label{newbound111} 
&&Q(\mathcal{E}_{\eta,N}) \leq Q({\mathcal{E}}^{e}_{\eta,N})=\log_2(\tfrac{\eta}{1-\eta})\\ \nonumber&&+h((1-\eta)(2N+1)+\eta)-h(\eta(2N+1)+1-\eta)\,.
\end{eqnarray} 
Once more this upper bound should be compared with previous upper bounds. At low noise, that is at low $N$ and high $\eta$, 
the best upper bound available is once more an upper bound for the generic two-way quantum capacity~\cite{Pirandola upp bound}:
\begin{equation}\label{PLOB11} 
Q(\mathcal{E}_{\eta,N})\leq Q_{\mathrm{PLOB}}(\eta,N)= -\log_2((1-\eta)\eta^{N})-h(2N+1)\;.
\end{equation}
Other bounds come from data processing \cite{Rosati,swat, Noh2018}, the best in the low noise regime being:
\begin{figure}[!tbp]
	\centering
	{\includegraphics[width=0.5\textwidth]{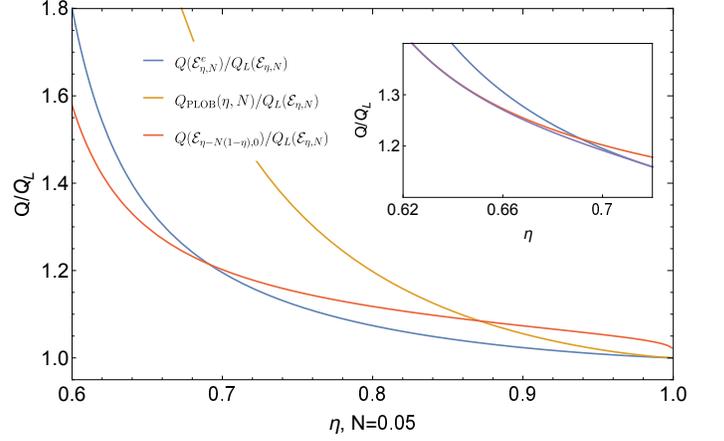}}
	\caption{Thermal attenuator: ratio between the upper bounds $Q(\mathcal{E}_{\eta-N(1-\eta),0})$~\cite{Rosati}, $Q_{\mathrm{PLOB}}(\eta,N)$~\cite{Pirandola upp bound} , $Q(\mathcal E^{e}_{\eta,N})$ (this work), and $Q_L(\mathcal{E}_{\eta,N})$ for $N=0.05$. {In the inset we plot a close-up in the region where $Q(\mathcal{E}_{\eta-N(1-\eta),0})$ and $Q(\mathcal E^{e}_{\eta,N})$ intersect. The purple line is the improved bound using the argument of Eq.~(\ref{betterdecoatt})}. }\label{fig:attcompratio}
\end{figure}
\begin{equation} \label{ROSATI} 
Q(\mathcal{E}_{\eta,N}) \leq Q(\mathcal{E}_{\eta-N(1-\eta),0})= \log_2\tfrac{\eta-N(1-\eta)}{(N+1)(1-\eta)}\;,
\end{equation}
A lower bound on $Q(\mathcal{E}_{\eta,N})$ is {given by the coherent information for one use of the channel, evaluated on an infinite temperature state,}
\begin{equation}
Q(\mathcal{E}_{\eta,N}) \geq Q_L(\mathcal{E}_{\eta,N}):=\max\{\log_2(\tfrac{\eta}{1-\eta})-h(2N+1),0\}.
\end{equation}
A comparison between all these curves  is reported in  Fig. \ref{fig:attcompratio}, showing that while our inequality~(\ref{newbound111}) performs worse than (\ref{ROSATI}) for low $\eta$, it
gives an improvement with respect to (\ref{PLOB11}) for high $\eta$. 
We finally remark that in our construction the choice of $\ketbra{0}_{A'}$ in the definition Eq.~(\ref{extatt}) of the extended attenuator is not necessarily optimal. Other Gaussian states could be chosen and the analysis could be done in the same way. In particular, the extension $\mathcal F_{\eta,N}\otimes \mathcal I[\rho_{A}\otimes\ketbra{\tau'}_{A'B}]$  gives a slightly better upper bound (optimizing over the single parameter in $\ket{\tau'}$ states), not noticeable on the plot in Fig. \ref{fig:attcompratio}.

\textit{Conclusion and Remarks}.--  The new bounds contained in this paper complement the bounds in~\cite{swat,Pirandola upp bound,Rosati,Noh2018}. They  have been determined  using degradable extensions and data-processing. This contribution extends the applicability of this technique, which now gives the best upper bounds at low noise for a large collection of channels of physical interest. In particular, the flagged  extension of the AGNC nicely generalizes the construction of flagged extensions to infinite dimensional channels, improving the upper bound on the quantum capacity by a considerable margin.\\
{The various bounds available are not directly comparable on the whole parameter region. However, taking into account all the possible data-processing decompositions
\begin{align}
\mathcal E_{\eta,N}=\mathcal E_{\eta_a,N_{a,1} }\circ\Phi_{g_a,N_{a,2}}=\Phi_{g'_{a},N'_{a,1}}\circ \mathcal E_{\eta'_{a},N'_{a,2} }\,,\label{betterdecoatt}\\
\Phi_{g,N}=\mathcal E_{\eta_{b},N_{b,1} }\circ\Phi_{g_{b},N_{b,2}}=\Phi_{g'_{b},N'_{b,1}}\circ \mathcal E_{\eta'_b,N'_{b,2}}\label{betterdecoamp}\,,
\end{align} 
using available direct bounds on the channels appearing in the decompositions, and minimizing over the decompositions, one can combine the bounds from the previous papers and the present ones, obtaining even better upper bounds. An example is seen in the inset of Fig.~\ref{fig:attcompratio}.
}

We mention also that the results presented here could be easily applied also for the case of energy constrained quantum capacity. 

\textit{Acknowledgements}.-- We thank Ludovico Lami, Matteo Rosati, and Xin Wang for helpful comments  and
 acknowledge support by MIUR via PRIN 2017 (Progetto di Ricerca di Interesse Nazionale): project QUSHIP (2017SRNBRK).

	\let\oldaddcontentsline\addcontentsline
\renewcommand{\addcontentsline}[3]{}

	\appendix
	\begin{widetext}
		\begin{center} \bf SUPPLEMENTAL MATERIAL\end{center}

	\section{Proof of Gaussianity of flagged additive Gaussian noise}
In this section we review the class of BGCs known as classical mixing channels, then show that flagged additive Gaussian noise is a classical mixing channel.

{ For any $n\times n$ square matrix $Y\geq 0$ with eigenvalues $\lambda_1,...,\lambda_n$, let us indicate the support of $Y$ as $S(Y)$, $\det_{+}Y=\prod_{i:\lambda_{i}>0}\lambda_i$, and pseudoinverse of $Y$ as  $Y^{\ominus 1}$. Classical mixing channels  have the form:
	\begin{equation}
	\Lambda_Y[\hat\rho]:=\int_{S(Y)}\mathrm{d} {\mathbf{r}}  \frac{e^{-\mathbf{r}^{\sf T}Y^{\ominus1}\mathbf{r}} }{\sqrt{\pi}^{\mathrm{dim}S(Y)} \sqrt{\mathrm{det_+} Y}}\hat D_{\mathbf{r}}\hat\rho \hat D_{\mathbf{r}}^\dagger\;.
	\end{equation} 
	Channels of this type are Gaussian and the action on the first and second moments is
	\begin{align}
	\mathbf{m}\xrightarrow{{\Lambda}_{Y}} \mathbf{m}'=\mathbf{m} \;, \qquad 
	V\xrightarrow{{\Lambda}_{Y}}V'=V + Y\;.
	\end{align}
	By direct comparison with Eq.~(16) of the main text it follows that 
	the flagged additive Gaussian noise  of  $\Lambda^{e}_{\beta}$ is 
	a classical mixing channel applied to the state $\hat\rho\otimes\ketbra{\beta/2}\otimes\ketbra{\beta/2}$ with the matrix $Y$ equal to}
\begin{equation}
Y={\left(\begin{array}{cccccc}
	\frac{2}{\beta} & 0 & 0 & 0 & 0 & -\frac{1}{\beta} \\
	0 & \frac{2}{\beta}& 0 & \frac{1}{\beta} & 0 & 0 \\
	0 & 0 & 0& 0 & 0 & 0 \\
	0 & \frac{1}{\beta} & 0 & \frac{1}{2\beta} & 0 & 0 \\
	0 & 0 & 0 & 0 & 0 & 0 \\
	-\frac{1}{\beta} & 0 & 0 & 0 & 0 & \frac{1}{2\beta} \\
	\end{array}\right)\;,}
\end{equation}
and thus Gaussian.
\section{Computing the Coherent information of extended channels}
In this section, we review how to compute the von Neumann entropy of Gaussian states, then we calculate the coherent information of the flagged additive Gaussian noise
	$\Lambda^{e}_{\beta}$ (see Eq.~(16) of the main text) and the extended thermal attenuator $\mathcal{E}^{e}_{\eta,N}$ (see Eq.~(27) of the main text).
The von~Neumann entropy of a Gaussian state can be computed from its covariance matrix $V$. In particular, we need to compute the symplectic eigenvalues ${d_1,...,d_n}$ of $V$, such that $V=SDS^{T}$ for some symplectic matrix $S$ and $D$ diagonal with elements $d_1,d_1,d_2,d_2,...,d_n, d_n$. In particular, these eigenvalues can be obtained from the eigenvalues of the matrix $i\Omega V$, which correspond to ${d_1,-d_1,...,d_n,-d_n}$. The Von~Neumann entropy of a state $\hat\rho$ with covariance matrix $V$ is then
\begin{equation}
S(\hat\rho)=\sum_{i=1}^{n} h(d_n)\;,
\end{equation}
with 
\begin{equation}
h(x):=\tfrac{x+1}{2}\log_2\left(\tfrac{x+1}{2}\right)-\tfrac{x-1}{2}\log_2\left(\tfrac{x-1}{2}\right)\;. 
\end{equation}
For flagged additive Gaussian noise channel $\Lambda^{e}_{\beta}$  and extended thermal attenuator $\mathcal{E}^{e}_{\eta,N}$, gaussian states maximize the coherent information since the channels are degradable and admit a Gaussian degrading map, satisfying the conditions of Theorem 12.40 of~\cite{HOL BOOK}. The degrading map is explicitly Gaussian for the extended thermal attenuator; in the case of the flagged additive noise the degrading map we present in the next section is not explicitly Gaussian, but its complementary is Gaussian. Since any Gaussian channel admit a Stinespring representation with a Gaussian unitary, there exists a Gaussian dilation of the complementary of the degrading map. By the properties of the Stinespring representation, the Stinespring dilation of the degrading map is isometric to a Gaussian isometry, with the connecting isometry acting trivially on the systems associated with the output of the complementary of the degrading map. This is enough to apply 
Theorem 12.40 of~\cite{HOL BOOK}. An explicit Gaussian degrading map will be presented elsewhere.

If in addition the channel is gauge-covariant, the maximization can be restricted to gauge-invariant states~\cite{HOL BOOK}. In our case, both the thermal attenuator and the flagged additive noise satisfy a generalized gauge-covariance property. Defining the Gaussian unitary on one mode $\hat R(\theta)$ acting on $\hat {\mathbf r}=(\hat x,\hat p)$ as the rotation matrix $R(\theta):=\left(\begin{matrix}\cos{\theta} & \sin{\theta}\\ -\sin{\theta} &\cos{\theta} \end{matrix}\right)$

\begin{equation}
\hat R(\theta)\hat {\mathbf r} \hat R(\theta)^{\dagger}=R(\theta)\hat{\mathbf r},
\end{equation}

we have

\begin{equation}\label{covadd}
\Lambda^{e}_{\beta}[\hat R(\theta)\hat \rho \hat R(\theta)^{\dagger}]=\hat R'(\theta)\Lambda^{e}_{\beta}[\hat \rho ]\hat R'(\theta)^{\dagger},
\end{equation}

with $\hat R'(\theta)$ being a three mode Gaussian unitary acting on $\hat {\mathbf r}=(\hat x_1,\hat p_1,\hat x_2,\hat p_2,\hat x_3,\hat p_3)$ as

\begin{equation}
\hat R'(\theta)\hat {\mathbf r} \hat R'(\theta)^{\dagger}=R'(\theta)\hat{\mathbf r}, 
\end{equation}

with 

\begin{align}
R'(\theta):=
\left(
\begin{array}{cccccc}
 \cos \theta & \sin \theta & 0 & 0 & 0 & 0 \\
 -\sin \theta & \cos \theta & 0 & 0 & 0 & 0 \\
 0 & 0 & \cos \theta & 0 & \sin \theta & 0 \\
 0 & 0 & 0 & \cos \theta & 0 & \sin \theta \\
 0 & 0 & -\sin \theta & 0 & \cos \theta & 0 \\
 0 & 0 & 0 & -\sin \theta & 0 & \cos \theta \\
\end{array}
\right).
\end{align}

In a similar way,

\begin{equation}\label{covtherm}
\mathcal{E}^{e}_{\eta,N}[\hat R(\theta)\hat \rho \hat R(\theta)^{\dagger}]=\hat R''(\theta)\mathcal{E}^{e}_{\eta,N}[\hat \rho ]\hat R''(\theta)^{\dagger},
\end{equation}

with $\hat R''(\theta)$ being a two mode Gaussian unitary acting on $\hat {\mathbf r}=(\hat x_1,\hat p_1,\hat x_2,\hat p_2)$ as

\begin{equation}
\hat R''(\theta)\hat {\mathbf r} \hat R''(\theta)^{\dagger}=R''(\theta)\hat{\mathbf r}, 
\end{equation}

with 

\begin{align}
R''(\theta):=
\left(
\begin{array}{cccc}
 \cos \theta & \sin \theta & 0 & 0  \\
 -\sin \theta & \cos \theta & 0 & 0 \\
 0 & 0 & \cos \theta &-\sin \theta  \\
 0 & 0 &  \sin \theta & \cos \theta \\
\end{array}
\right).
\end{align}

Adapting the argument in~\cite{HOL BOOK} for gauge-covariant channels, Eq.~(\ref{covadd}),(\ref{covtherm}), together with the concavity of the coherent information of degradable channels, imply that the maximum of the coherent information is attained on gauge-invariant Gaussian states, which coincides with thermal states for channels with one mode as input.  Moreover, since $\Lambda^{e}_{\beta}\circ \hat D_{\mathbf{s}}=\hat D_{\mathbf{s}}\circ \Lambda^{e}_{\beta}$ and $\mathcal{E}^{e}_{\eta,N}\hat D_{\mathbf{s}}=\hat D_{\sqrt{\eta}\mathbf{s}}\circ\mathcal{E}^{e}_{\eta,N}$, by concavity and unitarily invariance of the coherent information we have that higher energy thermal states have higher coherent information.

Thus, to compute the coherent information of the flagged additive noise we have to find the covariance matrix $V_M$ of $\Lambda^e_\beta[\hat\rho_{M}]$ and the covariance matrix $V'_M$ of  $(\Lambda^e_\beta\otimes \mathcal I) [\vert\rho_{M}\rangle\rangle\langle\langle\rho_{M}\vert]$ where $\hat\rho_{M}$ is the thermal state with average photon number $M$ and $\vert\rho_{M}\rangle\rangle$ is its purification, which can be taken to be the two-mode squeezed state
$\ket{\tau}$.

We obtain 
\begin{align}
&V_M=\left(
\begin{array}{cccccc}
2M+1+\frac{2}{\beta}& 0 & 0 & 0 & 0 & -\frac{1}{\beta} \\
0 & 2M+1+\frac{2}{\beta} & 0 & \frac{1}{\beta} & 0 & 0 \\
0 & 0 & \frac{2}{\beta} & 0 & 0 & 0 \\
0 & \frac{1}{\beta} & 0 & \frac{\beta}{2}+\frac{1}{2\beta} & 0 & 0 \\
0 & 0 & 0 & 0 & \frac{2}{\beta} & 0 \\
-\frac{1}{\beta} & 0 & 0 & 0 & 0 & \frac{\beta}{2}+\frac{1}{2\beta} \\
\end{array}
\right)\\
&V'_M=\left(
\begin{array}{cccccccc}
2M+1+\frac{2}{\beta} & 0 & 0 & 0 & 0 & -\frac{1}{\beta} & 2\sqrt{M(M+1)} & 0 \\
0 & 2 M+1+\frac{2}{\beta}  & 0 & \frac{1}{\beta} & 0 & 0 & 0 & -2\sqrt{M(M+1)} \\
0 & 0 & \frac{2}{\beta} & 0 & 0 & 0 & 0 & 0 \\
0 & \frac{1}{\beta} & 0 & \frac{\beta}{2}+\frac{1}{2\beta} & 0 & 0 & 0 & 0 \\
0 & 0 & 0 & 0 & \frac{2}{\beta} & 0 & 0 & 0 \\
-\frac{1}{\beta} & 0 & 0 & 0 & 0 & \frac{\beta}{2}+\frac{1}{2\beta} & 0 & 0 \\
2\sqrt{M(M+1)} & 0 & 0 & 0 & 0 & 0 & 2M +1 & 0 \\
0 & -2\sqrt{M(M+1)} & 0 & 0 & 0 & 0 & 0 & 2M+1 \\
\end{array}
\right)\, .
\end{align}
The eigenvalues of $i\Omega V_M$ are
\begin{equation}
\pm 2M+O(1), \qquad \pm \frac{\sqrt{1+\beta^2}}{\beta}+O(1/M) \;,\qquad \pm \frac{\sqrt{1+\beta^2}}{\beta}+O(1/M) \;,
\end{equation}
while the eigenvalues if $i\Omega V'_M$ are
\begin{equation}
\pm 2\frac{1}{\beta^{1/2}}\sqrt{M}+O(1), \qquad \pm 2\frac{1}{\beta^{1/2}}\sqrt{M}+O(1)\;, \qquad \pm 1\;, \qquad \pm 1 \;.
\end{equation}
{	Therefore we have
	\begin{equation}
	Q( \Lambda^e_\beta)=\lim_{M\rightarrow\infty}S(\Lambda^e_{\beta}[\rho_M])-S((\Lambda^e_{\beta}\otimes \mathcal{I})[\ketbra{\tau}{\tau}_M])=\log_2{\beta}-{1}/{\log 2}+2h \left(\tfrac{\sqrt{1+\beta^2}}{\beta}\right)\, ,
	\end{equation}
	as indicated in Eq.~(21) of the main text. }

To compute the coherent information of the extended thermal attenuator ${\mathcal{E}}_{\eta,N}^e$ we have to find the covariance matrix $V_M$ of ${\mathcal{E}}_{\eta,N}^e[\hat\rho_{M}]$ and the covariance matrix $V'_M$ of the complementary channel ${\mathcal{E}}_{\eta,N}^{e,c} [\hat\rho_{M}]={\mathcal{E}}_{1-\eta,N}^e [\hat\rho_{M}]$ where $\hat\rho_{M}$ is again the thermal state with average photon number $M$. 
We obtain 
\begin{align}
&V_M=\left(
\begin{array}{cccccccc}
\eta (2M+1) +(1-\eta)\eta (2N+1) & 0 & (1-\eta)2\sqrt{N(N+1)} & 0 \\
0 & \eta (2M+1) +(1-\eta)(2N+1)  & 0 & -(1-\eta)2\sqrt{N(N+1)} \\
(1-\eta)2\sqrt{N(N+1)} & 0 & \eta+(1-\eta)(2N+1)& 0 \\
0 & -(1-\eta)2\sqrt{N(N+1)} & 0 & \eta +(1-\eta)(2N+1) \\
\end{array}
\right)\, .
\end{align}
{while $V'_M$ is obtained from the above expression by exchanging $\eta\rightarrow 1-\eta$.}
The eigenvalues if $i\Omega V_M$ are hence
\begin{equation}
\pm \eta M+O(1), \qquad \pm (\eta+(1-\eta)(2N+1)) +O(1/M)\;, 
\end{equation}
while the eigenvalues if $i\Omega V'_M$ are
\begin{equation}
\pm (1-\eta) M+O(1), \qquad \pm ((1-\eta)+\eta(2N+1)) +O(1/M)\;.
\end{equation}
{Therefore we have
	\begin{eqnarray}
	Q({\mathcal E}_{\eta,N}^e)&=&\lim_{M\rightarrow\infty}S({\mathcal E}_{\eta,N}^e[\hat\rho_M])-S({\mathcal E}_{1-\eta,N}^e[\hat\rho_M]) \nonumber \\
	\nonumber &=&-\log_2\left(\tfrac{\eta}{1-\eta}\right)+h(\eta+(1-\eta)(2N+1))-h((1-\eta)+\eta(2N+1))\;,
	\end{eqnarray}
	as indicated in Eq.~(32) of the main text.}

\section{Degradability of flagged additive noise channel}\label{app:degraddnoise}

Here we prove the degradability of the channel $\Lambda^{e}_{\beta}$. We define the unitary operator $\hat U^{(x)}:L_2(\mathbb{R}^2)\rightarrow L_2(\mathbb{R}^2)$
\begin{equation}
\hat{U}^{(x)}:\psi(x_1,x_2)\rightarrow \psi(x_1,x_2+x_1)\;, 
\end{equation}
and $\hat U^{(p)}:L_2(\mathbb{R}^2)\rightarrow L_2(\mathbb{R}^2)$
\begin{equation}
\hat U^{(p)}:\psi(x_1,x_2)\rightarrow \psi(x_1,x_2)e^{-i x_1 x_2} \;. 
\end{equation}
We define the pure states of two modes $\ket{C_1}$ and $\ket{C_2}$ with wave functions respectively
\begin{align}
\psi_{x,\beta}(x_1,x_2)&=\sqrt{\frac{\beta}{2\pi}}e^{-\beta\frac{ x_1^2}{4}-i\frac{x_1x_2}{2}-\beta \frac{x_2^2}{4}}\;, \nonumber\\ \psi_{p,\beta}(x_1,x_2)&=\sqrt{\frac{\beta}{2\pi}}e^{-\beta\frac{ x_1^2}{4}+i\frac{x_1x_2}{2}-\beta \frac{x_2^2}{4}}\;.
\end{align}
We also note that the wave function of a displaced squeezed state $\hat D_{(0,x)}\ket{\beta}$ with $x\in \mathbb R$ is
\begin{equation}
\psi_{\beta,x}(x')=\sqrt{\frac{\beta}{\pi }}e^{-\beta\frac{x'^2}{2}-ix'x}\;. 
\end{equation}
Note hence that 
\begin{align}
&\bra{\psi}_{A}\otimes \bra{C_1}_{XP'}\otimes \bra{C_2}_{PX'}{\hat{U}^{(x)}_{XA}}{}^{\dagger}{\hat{U}^{(p)}_{PA}}{}^{\dagger}[f(x_{X},x_{P})]=\nonumber\\
&=\sqrt{\frac{\beta}{2\pi}}\int_{\mathbb R^2} \mathrm {d} x_{X'}\mathrm{d} x_{P'} \bra{\psi}{\hat D}{}^{\dagger}_{(x_{X}, 0)}{\hat D}{}^{\dagger}_{(0,x_{P})}\bra{\beta/2}_{P'}{\hat D}{}^{\dagger}_{(0,x_{X}/2)}\bra{\beta/2}_{X'}\hat{D}{}^{\dagger}_{(0,- x_{P}/2)} e^{-\frac{\beta}{4}( x_{X}^2+x_{P}^2)}f(x_{X},x_{P})\;. 
\end{align}
It follows that
\begin{align}
&\Lambda^{e}_{\beta}[\ketbra{\psi}]:=\Tr_{X P}[\hat{U}^{(p)}_{PA}\hat{U}^{(x)}_{XA}\ketbra{\psi}_{A}\otimes\ketbra{C_1}_{XP'}\otimes \ketbra{C_2}_{PX'}{U^{(x)}_{XA}}^{\dagger}{U^{(p)}_{PA}}^{\dagger}]\nonumber\\
&=\frac{\beta}{2\pi} \int_{\mathbb R^2} \mathrm {d} x_{X}\mathrm{d} x_{P} e^{-\frac{\beta}{2}( x_{X}^2+x_{P}^2)}\hat D_{(x_{X},x_{P})}\hat \rho \hat D_{(x_{X},x_{P})}^{\dagger}\nonumber\\&\otimes\nonumber\hat D_{(0,-x_{P}/2)}\ketbra{\beta/2}_{X'}\hat D_{(0,-x_{P}/2)}^{\dagger}\otimes \hat D_{(0,x_{X}/2)}\ketbra{\beta/2}_{P'} \hat D_{(0,x_{X}/2)}^{\dagger}\;.
\end{align}
The wave function of $U^{(p)}_{P'A}U^{(x)}_{X'A}U^{(p)}_{PA}U^{(x)}_{XA}\ket{\psi}_{A}\otimes \ket{C_1}_{XP'}\otimes \ket{C_2}_{PX'}$ is
\begin{align}
&\sqrt{\frac{\beta}{2\pi}}e^{-\beta \frac{x_X^2}{4}-i\frac{x_Xx_{P'}}{2}-\beta \frac{x_{P'}^2}{4}}e^{-\beta \frac{x_P^2}{4}+i\frac{x_Px_{X'}}{2}-\beta \frac{x_{X'}^2}{4}}\psi(x_A+x_X+x_{X'})e^{i(x_A-x_{X'})x_{P}+ i x_{A}x_{P'}}\\
&=\sqrt{\frac{\beta}{2\pi}}e^{-\beta \frac{x_X^2}{4}-i\frac{x_Xx_{P'}}{2}-\beta \frac{x_{X'}^2}{4}}e^{-\beta \frac{x_P^2}{4}-i\frac{x_Px_{X'}}{2}-\beta \frac{x_{P'}^2}{4}}\psi(x_A+x_X+x_{X'})e^{i x_A(x_{P}+x_{P'})}\;.
\end{align}
Since this wave function is symmetric under exchange $X\leftrightarrow X'$, $P\leftrightarrow P'$, defining the map
\begin{equation}
W_{X'P'A\rightarrow X' P'}[\hat \rho]:=\Tr_{A}[\hat{U}^{(p)}_{P'A}\hat{U}^{(x)}_{X'A}\hat{\rho}_{X'P'A}{\hat{U}^{(x)}_{X'A}}{}^{\dagger}{\hat{U}^{(p)}_{P'A}}{}^\dagger]\;,
\end{equation}
we have that
\begin{align}
&W_{X'P'A\rightarrow X' P'}\circ {\Lambda_{\beta}}_{A\rightarrow X'P'A}[\rho]=\Tr_{XPA}[\hat{U}^{(p)}_{P'A}\hat{U}^{(x)}_{X'A}\hat{U}^{(p)}_{P'A}\hat{U}^{(x)}_{X'A}\hat \rho_{A}\otimes \ketbra{C_1}_{XP'}\otimes \ketbra{C_2}_{PX'}{(\hat{U}^{(p)}_{P'A}\hat{U}^{(x)}_{X'A}\hat{U}^{(p)}_{P'A}\hat{U}^{(x)}_{X'A})}{}^{\dagger}]\nonumber\\
&=\Tr_{X'P'A}[\hat{U}^{(p)}_{P'A}\hat{U}^{(x)}_{X'A}\hat{U}^{(p)}_{P'A}\hat{U}^{(x)}_{X'A}\hat \rho_{A}\otimes \ketbra{C_1}_{XP'}\otimes \ketbra{C_2}_{PX'}{(\hat{U}^{(p)}_{P'A}\hat{U}^{(x)}_{X'A}\hat{U}^{(p)}_{P'A}\hat{U}^{(x)}_{X'A})}^{\dagger}]={\Lambda_{\beta}^{e,c}}_{A\rightarrow X P}[\hat \rho_{A}]\;.
\end{align}
Thus $\Lambda_{\beta}^{e,c}$ is degradable.

\end{widetext}
\end{document}